# Recurrent computations for visual pattern completion

Hanlin Tang[1,4*], Martin Schrimpf[2,4*], Bill Lotter[1,3,4*], Charlotte Moerman[4], Ana Paredes[4], Josue Ortega Caro[4], Walter Hardesty[4], David Cox[3], Gabriel Kreiman[4] ✎

*These authors contributed equally
✎To whom correspondence should be addressed at
gabriel.kreiman@tch.harvard.edu

[1]Program in Biophysics, Harvard University
[2]Program in Software Engineering, Augsburg University, Technische Universität München, Ludwig-Maximilians-Universität München
[3]Molecular and Cellular Biology, Harvard University
[4]Children's Hospital, Harvard Medical School

Text Statistics:
Number of words in abstract: 164
Number of words in significance statement: 100
Number of Figures: 4
Number of Tables: 0
Number of Supplementary Figures: 10

**Abstract**

Making inferences from partial information constitutes a critical aspect of cognition. During visual perception, pattern completion enables recognition of poorly visible or occluded objects. We combined psychophysics, physiology and computational models to test the hypothesis that pattern completion is implemented by recurrent computations and present three pieces of evidence that are consistent with this hypothesis. First, subjects robustly recognized objects even when rendered <15% visible, but recognition was largely impaired when processing was interrupted by backward masking. Second, invasive physiological responses along the human ventral cortex exhibited visually selective responses to partially visible objects that were delayed compared to whole objects, suggesting the need for additional computations. These physiological delays were correlated with the effects of backward masking. Third, state-of-the-art feed-forward computational architectures were not robust to partial visibility. However, recognition performance was recovered when the model was augmented with attractor-based recurrent connectivity. These results provide a strong argument of plausibility for the role of recurrent computations in making visual inferences from partial information.

**Significance Statement**

The ability to complete patterns and interpret partial information is a central property of intelligence. Deep convolutional network architectures have proved successful in labeling whole objects in images and capturing the initial 150 ms of processing along the ventral visual cortex. This study shows that human object recognition abilities remain robust when only small amounts of information are available due to heavy occlusion, but the performance of bottom-up computational models is impaired under limited visibility. This study provides behavioral, neurophysiological and computational evidence suggesting that recurrent computations may help the brain solve the fundamental challenge of pattern completion.

\body

Humans and other animals have a remarkable ability to make inferences from partial data across all cognitive domains. This inference capacity is ubiquitously illustrated during pattern completion to recognize objects that are partially visible due to noise, limited viewing angles, poor illumination or occlusion. There has been significant progress in describing the neural machinery along the ventral visual stream responsible for recognizing whole objects (1-5). Computational models instantiating biologically plausible algorithms for pattern recognition of whole objects typically consist of a sequence of filtering and non-linear pooling operations. The concatenation of these operations transforms pixel inputs into a feature representation amenable for linear decoding of object labels. Such feed-forward algorithms perform well in large-scale computer vision experiments for pattern recognition (6-10) and provide a first-order approximation to describe the activity of cortical neurons (e.g., (11)).

Spatial and temporal integration play an important role in pattern completion mechanisms (12-15). When an object is occluded, there are infinitely many possible contours that could join the object's parts together. Yet, the brain typically manages to integrate those parts to correctly recognize the occluded object. Multiple studies have highlighted the importance of temporal integration by demonstrating that recognizing partially visible objects takes more time than recognizing fully visible ones at the behavioral (12, 16) and physiological levels (13, 14, 17). We conjectured that within-layer and top-down recurrent computations are involved in implementing the spatial and temporal integrative mechanisms underlying pattern completion. Recurrent connections can link signals over space within a layer and provide specific top-down modulation from neurons with larger receptive fields (18, 19). Additionally, recurrent signals temporally lag their feed-forward counterparts and therefore provide an ideal way to incorporate temporal integration mechanisms.

To examine plausible mechanisms involved in pattern completion, we combined psychophysics, neurophysiology (14), and computational modeling to evaluate recognition of partially visible objects. We show that humans robustly recognize objects even from limited amount of information, but performance rapidly deteriorates when computations are interrupted by a noise mask. On an image-by-image basis, the behavioral effect of such backward masking correlates with an increase in latency in neurophysiological intracranial field potentials along the ventral visual stream. A family of modern feed-forward convolutional hierarchical models is *not* robust to occlusion. We extend previous notions of attractor dynamics by adding recurrence to such bottom-up models and providing a proof-of-concept model that captures the essence of human pattern completion behavior.

## Results

**Robust recognition of partially visible objects**. Subjects performed a recognition task (**Fig. 1A-B**) involving categorization of objects that were either partially visible ("Partial", **Fig. 1C** right) or fully visible ("Whole", **Fig. 1C** left). Images were followed either by a gray screen ("unmasked", **Fig. 1A**) or a spatially overlapping noise pattern ("masked", **Fig. 1B**). The image presentation time, referred to as stimulus onset asynchrony (SOA), ranged from 25 to 150 ms in randomly ordered trials. Stimuli consisted of 325 objects belonging to 5 categories: animals, chairs, faces, fruits, and vehicles. The parts revealed for each object were chosen randomly. There were 40 images per object, comprising a total of 13,000 images of partial objects (**Methods**).

For whole objects and without a mask, behavioral performance was near ceiling, as expected (**Fig. 1F**, 100% visible). Subjects robustly recognized partial objects across a wide range of visibility levels despite the limited information provided (**Fig. 1F**). Although poor visibility degraded performance, subjects still showed 80±3% performance at 35±2.5% visibility (partial versus whole objects: $p<10^{-10}$, two-sided t-test). Even for images with 10±2.5% visibility, performance was well above chance levels (59±2%, $p<10^{-10}$, two-sided t-test, chance = 20%). There was a small but significant improvement in performance at longer SOAs for partially visible objects (**Fig. 1H** dashed lines, Pearson *r*, = 0.56, p<0.001, permutation test).

In a separate experiment, we generated images where objects appeared behind a black surface occluder (**Fig. 1D**). Consistent with previous studies (e.g. (15)), recognition was also robust when using heavily occluded images (**Fig. 1I**). The presence of an occluder improved recognition performance with respect to partial objects (compare **Fig. S1A** versus **S1B,** $p<10^{-4}$, Chi-squared test). We focused next on the essential aspects of pattern completion by considering the more challenging condition of partially visible objects, without help from other cues such as occluders.

While subjects had not seen any of the *specific* images in this experiment before, they had had extensive experience with fully visible and occluded versions of *other* images of animals, faces, fruits, chairs and vehicles. We conducted a separate experiment with novel shapes (**Fig. 1E**, **Fig. S8A**) to assess whether robustness to limited visibility (**Fig. 1F, H, I**) extended to unfamiliar objects. Visual categorization of such novel objects was also robust to limited visibility (**Fig. 1J**, **Fig. S8B**).

**Backward masking disrupts recognition of partially visible objects**. Behavioral (20), physiological (21, 22), and computational studies (3, 4, 11) suggest that recognition of whole isolated objects can be described by rapid, largely feed-forward, mechanisms. Several investigators have used backward masking to force visual recognition to operate in a fast regime with minimal influences from recurrent signals (23): when an image is rapidly followed by a spatially overlapping mask, the high-contrast noise mask interrupts any additional, presumably recurrent, processing of the original image (24-26). We asked whether this fast, essentially feed-forward, recognition regime imposed by backward masking is sufficient for robust recognition of partially visible objects by randomly interleaving trials with a mask (**Fig. 1B**).

Performance for whole images was affected by the mask only for the shortest SOA values (cf. **Fig. 1F** versus **1G** at 100% visibility, $p<0.01$, two-sided t-test). When partial objects were followed by a backward mask, performance was severely impaired (cf. **Fig. 1F** versus **1G**). A two-way ANOVA on performance with SOA and masking as factors revealed a significant interaction ($p<10^{-8}$). The behavioral consequences of shortening SOA were significantly stronger in the presence of backward masking (cf. solid versus dashed lines in **Fig. 1H**). Additionally, backward masking disrupted performance across a wide range of visibility levels for SOAs ≤ 100 ms (**Fig. 1G-H**). Similar effects of backward masking were observed when using occluded objects (**Fig. 1I,** $p<0.001$, two-way ANOVA) as well as when using novel objects (**Fig. 1J**, **Fig. S8C-D**, $p<0.0001$, two-way ANOVA). In sum, interrupting processing via backward masking led to a large reduction in the ability for recognition of partially visible objects, occluded images, and partially visible novel objects, across a wide range of SOA values and visibility levels.

**Images more susceptible to backward masking elicited longer neural delays along human ventral visual cortex.** In a recent study, we recorded invasive physiological signals throughout the ventral visual stream in human patients with epilepsy while they performed an experiment similar to the one in **Fig. 1A** (14). This experiment included 25 objects presented for 150 ms without any masking, with random bubble positions in each trial. For whole objects, neural signals along the ventral visual stream showed rapid selective responses to different categories, as shown for an example electrode in the left fusiform gyrus in **Fig. 2A-B**. When presenting partially visible objects, the neural signals remained visually selective (**Fig. 2C-D**). The visually selective signals elicited by the partial objects were significantly delayed with respect to the responses to whole objects (compare the neural latency defined here as the single trial time of peak responses in **Fig. 2C-D** with the time of peak response before 200 ms in **Fig. 2A-B**). Because the visible features varied from trial to trial, different renderings of the same object elicited a wide distribution in the neural latencies (**Fig. 2C-D**). For example, the peak voltage occurred at 206 ms post stimulus onset in response to the first image and at 248 ms in response to the last image in **Fig. 2C**.

Heterogeneity across different renderings of the same object was also evident in the range of effects of backward masking at the behavioral level in the experiment in **Fig. 1G-H**. We hypothesized that those images that elicited longer neural delays would also be more susceptible to backward masking. To test this hypothesis, we selected two electrodes in the neurophysiological study showing strong visually selective signals (**Methods**, one of these sites is shown in **Fig. 2A-D**). We considered 650 images of partially visible objects corresponding to the 25 objects from the neurophysiology experiment. Using the same images (i.e. the exact same features revealed for each object), we conducted a separate psychophysics experiment to evaluate the effect of backward masking on each individual image (n=33 subjects). This experiment allowed us to construct a curve of behavioral performance as a function of SOA during backward masking, for each of the selected images from the neurophysiology experiment (**Fig. 2E**). To quantify the effect of backward masking for each individual image, we defined a masking index, MI = 1-AUC, where AUC is the normalized area under the curve in the performance versus SOA plot (gray area in **Fig. 2E**). Larger MI values correspond to larger effects of backward masking: MI ranges from 0 (no effect of backward masking) to 0.8 (backward masking leads to chance performance). For example, in **Fig. 2C**, the first image was less affected by backward masking than the last image, particularly at short SOA values (MI values of 3% and 20% respectively).

For those images from the preferred category for each of the two electrodes, the masking index showed a weak but significant correlation with the neural response latency, even after accounting for image difficulty and recording site differences (**Fig. 2F**, Pearson $r$ = 0.37, p = 0.004, permutation test, **Methods**). This effect was stimulus selective: the masking index was *not* correlated with the neural response latency for images from the non-preferred categories (p = 0.33, permutation test). The neural latencies are noisy measures based on single trials (**Fig. 2C**, **Methods**), the physiology and behavioral experiments were conducted in different subjects, and there was variability across subjects in the masking index (**Fig. 2F**, **S2**). Yet, despite all of these sources of noise, images that led to longer neural response latencies were associated with a stronger effect of interrupting computations via backward masking.

**Standard feed-forward models are not robust to occlusion**. We next investigated the potential computational mechanisms responsible for the behavioral and physiological observations in **Figs. 1-2**. We began by considering state-of-the-art implementations of purely feed-forward computational models of visual recognition. These computational models are characterized by hierarchical, feed-forward processing with progressive increases in the size of receptive fields, the degree of selectivity, and tolerance to object transformations (e.g. (2-4)). Such models have been successfully used to describe rapid recognition of whole objects at the behavioral level (e.g. (4)) and neuronal firing rates in areas V4 and inferior temporal cortex in macaque monkeys (e.g. (11)). Additionally, these deep convolutional network architectures achieve high performance in computer vision competitions evaluating object recognition capabilities (e.g. (6-8)).

We evaluated the performance of these feed-forward models in recognition of partially visible objects using the same 325 objects (13,000 trials) in **Fig. 1**. As a representative of this family of models, we considered AlexNet (6), an eight-layer convolutional neural network trained via back propagation on ImageNet, a large corpus of natural images (10). We used as features either activity in the last fully connected layer before readout (fc7, 4096 units), or activity in the last retinotopic layer (pool5, 9216 units). To measure the effect of low-level differences between categories (contrast, object area, etc.), we also considered raw pixels as baseline performance (256x256=65,636 features).

We sought to measure the robustness of these networks to partial object visibility in the same way that tolerance to other transformations such as size and position changes is evaluated, i.e., by training a decision boundary on one condition (specific size, viewpoint, whole objects) and testing on the other conditions (other sizes, viewpoints, occlusion; e.g. (2, 4)). It is not fair to compare models trained with occluded objects versus models trained exclusively with whole objects, and therefore we do not include occluded objects in the training set. Furthermore, the results in **Fig. 1J** and **S8** show that humans can perform pattern completion for novel objects without any prior training with occluded versions of those objects. We trained a support vector machine (SVM) classifier (linear kernel) on the features of *whole* objects, and tested object categorization performance on the representation of images of *partial* objects. Importantly, all the models were trained exclusively with whole objects and performance was evaluated in images with partially visible objects. Cross-validation was performed over objects: objects used to train the decision boundary did not appear as partial objects in the test set. The performance of raw pixels was essentially at chance level (**Fig. 3A**). In contrast, the other models performed well above chance (p<$10^{-10}$, two-sided t-test, see also **Fig. S4**). While feed-forward models performed well above chance, there was a significant gap with respect to human performance, at all visibility levels below 40% (p<0.001, Chi-squared test, **Fig. 3A**). These results are consistent with those reported in other simulations with occluded objects and similar networks (27). The decrease in performance of feed-forward models compared to humans depends strongly on the stimuli and on the amount of information available: bottom-up models were comparable to humans at full visibility (**Fig. S3A**, (28)).

The decline in performance with low visibility was not specific to the set of images used in this study: AlexNet pool5 and fc7 also performed below human levels when considering novel objects (**Fig. S9A**). The decline in performance with low visibility was not specific to using pixels, AlexNet pool5 or fc7 layers. All the feed-forward models that we tested led to the same conclusions, including different layers of VGG16, VGG19 (7), InceptionV3 (9), and ResNet50 (8) (**Fig. S4**). Among these models, the VGG16 architecture provided slightly better recognition performance in the low visibility regime.

The models shown in **Fig. 3A** and **S4** were trained to optimize object classification performance in the ImageNet 2012 data set (10) without any specific training for the set of objects used in our study, except for the SVM classifier. To assess whether fine-tuning the model's weights could alleviate the challenges with limited visibility, we fine-trained AlexNet via back-propagation using the 325 whole objects and then re-tested this fine-tuned model on the images with limited visibility. Fine tuning the AlexNet architecture did not lead to improvements at low visibility (**Fig. S5**). These results are consistent with a previous computational study, using feed-forward models similar to the ones in the current work and evaluating a more extensive image data set (27).

We used stochastic neighborhood embedding (t-SNE) to project the AlexNet fc7 layer features onto 2D and visualize the effects of occlusion on the model (**Fig. S3B**). The representation of whole objects (open circles) showed a clear separation among categories but partial objects from different categories (filled circles) were more similar to each other than to their whole object counterparts. Therefore, decision boundaries trained on whole objects did not generalize to categorization of partial objects (**Fig. 3A**). Despite the success of purely feed-forward models in recognition of whole objects, these models were not robust under limited visibility.

We next sought to further understand the breakdown in the models' representations of objects under partial visibility. Removing large amounts of pixels from the objects pushed the model's representation of the partially visible object away from their whole counterparts (e.g., arrow in **Fig. S3B**). The distance between the representation of a partially visible object and the corresponding whole object category mean is indicative of the impact of partial visibility. We evaluated whether this distortion was correlated with the latencies in the neural recordings from **Fig. 2**. We reasoned that images of partial objects whose model representation was more distorted would lead to longer neural response latencies. We computed the Euclidean distance between the representation of each partial object and the whole object category mean. We found a modest but significant correlation at the object-by-object level between the computational distance to the category mean and the neural response latency for the pool5 (**Fig. 3B**) and fc7 (**Fig. 3C**) features. The statistical significance of these correlations was assessed by regressing the distance to category mean against the neural latency, along with the following additional predictors to account for potential confounds: (i) the percentage of object visibility and pixel distance to regress out any variation explained by low-level effects of occlusion and difficulty; (ii) the electrode number to account for the inter-electrode variability in our dataset, and (iii) the masking index (**Fig. 2E**), to control for overall recognition difficulty. The model distance to category mean in the pool5 and fc7 layers correlated with the response latency beyond what could be explained by these additional factors (pool 5: Pearson $r$ = 0.27, $p$=0.004, permutation test; fc7: Pearson $r$ = 0.3, $p$=0.001, permutation test). In sum, state-of-the-art feed-forward architectures did not robustly extrapolate from whole to partially visible objects and failed to reach human-level performance in recognition of partially visible objects. As the difference in the representation of whole and partial objects increased, the time it took for a selective neural response to evolve for the partial objects was longer.

**Recurrent neural networks improve recognition of partially visible objects.** The behavioral, neural and modeling results presented above suggest a need for additional computational steps beyond those present in feed-forward architectures to build a robust representation for partially visible objects. Several computational ideas, originating from models proposed by Hopfield (29) have shown that attractor networks can perform pattern completion. In the Hopfield network, units are connected in an all-to-all fashion with weights defining fixed attractor points dictated by the whole objects to be represented. Images that are pushed farther away by limited visibility would require more processing time to converge to the appropriate attractor, consistent with the behavioral and physiological observations. As a proof-of-principle, we augmented the feed-forward models discussed in the previous section with recurrent connections to generate a robust representation through an attractor-like mechanism (**Fig. 4A**), with one attractor for each whole object. We used the AlexNet architecture with fixed feed-forward weights from pre-training on ImageNet (as in **Fig. 3**) and added recurrent connections to the fc7 layer. Recurrent connectivity is ubiquitous throughout *all* visual neocortical areas in biological systems. The motivation to include recurrent connectivity *only* in the fc7 layer was to examine first a simple and possibly minimal extension to the existing architectures (**Discussion**).

We denote the activity of the fc7 layer at time *t* as the 4096-dimensional feature vector $\mathbf{h}_t$. At each time step, $\mathbf{h}_t$ was determined by a combination of the activity from the previous time step $\mathbf{h}_{t-1}$ and the constant input from the previous layer $\mathbf{x}$: $\mathbf{h}_t = f(\mathbf{W}_h \mathbf{h}_{t-1}, \mathbf{x})$ where *f* introduces a non-linearity (**Methods**). The input from the previous layer, fc6, was kept constant and identical to that in the feed-forward AlexNet. $\mathbf{W}_h$ is a weight matrix that governs the temporal evolution of the fc7 layer. We considered a Hopfield network, RNN$_h$, without introducing any free parameters that depended on the partial objects, where $\mathbf{W}_h$ was a symmetric weight matrix dictated by the fc7 representation of the whole objects, using the implementation in (30). The initial state of the network was given by the activity in the previous layer, $\mathbf{h}_0 = \mathbf{W}_{6 \rightarrow 7} fc6$, followed by binarization. The state of the network evolved over time according to $\mathbf{h}_t = satlins(\mathbf{W}_h \mathbf{h}_{t-1})$ where *satlins* is a saturating non-linearity (**Methods**). We verified that the whole objects constituted an attractor point in the network by ensuring that their representation did not change over time when used as inputs to the model. We next evaluated the responses of RNN$_h$ to all the images containing partial objects. The model was run until convergence (i.e. until none of the feature signs changed between consecutive time steps). Based on the final time point, we evaluated the performance in recognizing partially visible objects. The RNN$_h$ demonstrated a significant improvement over the AlexNet fc7 layer (**Fig. 4B**, 57±0.4%, p<0.001, Chi-squared test).

The dynamic trajectory of the representation of whole and partial objects in the fc7 layer of the RNN$_h$ model is visualized in **Fig. 4C**. Before any recurrent computations have taken place, at t=0 (left), the representations of partial objects were clustered together (closed circles) and separated from the clusters of whole objects in each category (open circles) (**Fig. S3B**). As time progressed, the cluster of partial objects was pulled apart and moved towards their respective categories. For example, at t=16 (center) and t=256 (right), the representation of partial chairs (closed blue circles) largely overlapped with the cluster of whole chairs (open blue circles). Concomitant with this dynamic transformation in the representation of partial objects, the overall performance of the RNN$_h$ model improved over time (**Fig. 4D**).

In addition to the average performance reported in **Fig. 4B** and **4D**, we directly compared performance at the object-by-object level between humans and RNN$_h$ (**Fig. S6**). There were notable differences across categories (e.g. humans were much better than the model in detecting faces, green circles in **Fig. S6**). For this reason, we first compared models and humans at the object-by-object level within each category and then averaged the results across categories. Over time, RNN$_h$ behaved more like humans at the object-by-object level (**Fig. 4E**). For each time step in the model, we computed the average correct rate on partial objects for each object, from each of the 5 categories, and correlated this vector with the pattern of human performance (**Fig. S6**). The upper bound (dashed line in **Fig. 4E**) represents human-human similarity, defined as the correlation in the response patterns between half of the subject pool and the other half. Over time, the recurrent model-human correlation increased towards the human-human upper bound.

Adding a Hopfield-like recurrent architecture to AlexNet also improved performance in recognition of the novel objects illustrated in **Fig. S8A** (**Fig. S9B-D**). Similar conclusions were obtained when considering the VGG16 architecture and adding Hopfield-like recurrent connections to the fc1 layer (**Fig. S7**).

In sum, implementing recurrent connections in an attractor-like fashion at the top of a feed-forward hierarchical model significantly improved the model's performance in pattern completion, and the additional computations were consistent with temporal delays described at the behavioral and neural levels.

**Backward masking impaired RNN model performance**. We reasoned that the backward mask introduced in the experiment discussed in **Fig. 1B, G-H** impaired performance by interrupting processing and we set out to investigate whether we could reproduce this effect in the RNN$_h$ model. We computed the responses of the AlexNet model to the mask and fed the fc6 features for the mask to the RNN$_h$ model after a certain number of time steps. Switching the mask on at earlier time points was meant to mimic shorter SOA's in the psychophysical experiments. We read out performance based on the resulting fc7 activity combining the partial object and the mask at different time points (**Fig. 4F**). Presenting the mask reduced performance from 58±2% (SOA=256 time steps) to 37±2% (SOA=2 time steps). Although we cannot directly compare SOAs in milliseconds to time steps in the model, these results are qualitatively consistent with the behavioral effects of backward masking

(**Fig. 1H;** see side-by-side comparison of the physiological, behavioral and computational dynamics in **Fig. S10)**.

**Discussion**

It is routinely necessary to recognize objects that are partially visible due to occlusion and poor illumination. The visual system is capable of making inferences under such conditions even when only 10-20% of the object is visible (**Fig. 1F**), even for novel objects (**Fig. 1J**). We investigated the mechanisms underlying such robust recognition of partially visible objects (referred to as pattern completion) at the behavioral, physiological and computational levels. Backward masking impaired recognition of briefly presented partial images (25ms ≤ SOA ≤ 100 ms) (**Figs. 1G-J**). The strength of the disruptive effect of backward masking was correlated with the neural delays described previously from invasive recordings along the ventral visual stream (14) (**Fig. 2**). State-of-the-art bottom-up computational architectures trained on whole objects failed to achieve robustness in recognition of partially visible objects (**Fig. 3A, S4, S5**). The introduction of recurrent connections to the top layer led to significant improvement in recognition of objects from partial information at the average level (**Fig. 4B, S7, S9B**) and at the object-by-object level (**Fig. 4E, S6**). The increase in performance involved recurrent computations evolving over time that were interrupted by the introduction of a mask (**Fig. 4F**).

Recognition of partially visible objects requires longer reaction times (12, 16) and their neural representation is delayed with respect to that of whole objects (13, 14, 17). These delays suggest the need for additional computations to interpret partially visible images. Interrupting those additional computations by a backward mask significantly impairs recognition (**Fig. 1G-J**). Backward masking disproportionately affects recurrent computations (24-26). Accordingly, we conjectured that the disruptive effect of backward masking during pattern completion could be ascribed to the impairment of such recurrent computations. The rapid and selective signals along the ventral visual stream, which enable recognition of whole objects within ~150 ms reflect largely bottom-up processing (2-4, 11, 20-22). Physiological delays of about 50 ms during recognition of partial objects (13, 14) provide ample time for recurrent connections to exert their effects during pattern completion. These delays could involve recruitment of lateral connections (18) and/or top-down signals from higher visual areas (31).

We presented a proof-of-principle model that extended bottom-up architectures by adding recurrent connections at the top level (**Fig. 4A**). This extension improved performance (**Fig. 4B, S7, S9B**), showed a correlation at the object-by-object level with human recognition (**Fig. 4E**) and accounted for the effects of backward masking (**Fig. 4F, S10**). The RNN$_h$ model had no free parameters that depended on the partial objects: all the weights were determined by the whole objects.

Humans are constantly exposed to partially visible objects. While subjects had not previously seen the specific objects that we used, they had had experience with occluded animals, chairs, faces, fruits and vehicles. To evaluate whether category-specific experience with occluded objects is required for pattern completion, we conducted an experiment with completely novel objects (**Fig. 1J**, **S8-S9**). Subjects robustly categorized novel objects under low visibility even when they had never seen those heavily occluded objects or similar ones before.

There exist infinitely many possible bottom-up models. Even though we examined multiple state-of-the-art models that are successful in object recognition (AlexNet, VGG16, VGG19, InceptionV3, ResNet50), their failure to account for the behavioral and physiological results (**Fig. 3**, **S4**, (27, 28)) should be interpreted with caution. We do not imply that it is impossible for *any* bottom-up architecture to recognize partially visible objects. In fact, a recurrent network with a finite number of time steps can be unfolded into a bottom-up model by creating an additional layer for each time step. However, there are several advantages to recurrent architectures including a reduction in the number of units and weights. Furthermore, such unfolding of time into layers is only applicable when we know *a priori* the fixed number of computational steps, whereas recurrent architectures allow an arbitrary and dynamically flexible number of computations.

The RNN dynamics involve temporal evolution of the features (**Fig. 4C-F**), bringing the representation of partial objects closer to that of whole objects. These computational dynamics, map onto the temporal lags observed at the behavioral and physiological levels. Furthermore, these dynamics are interrupted by the presentation of a backward mask in close temporal proximity to the image (**Fig. 1G-J**, **4F, S10**).

Multiple other cues can aid recognition of partially visible objects including understanding relative positions, segmentation, movement, source of illumination, and stereopsis. Additionally, when children learn to recognize objects, they often encounter partially visible objects that they can explore in a continuous fashion from multiple different angles. It will be interesting to integrate these additional sources of information to understand how they contribute to the mechanisms of pattern completion. The convergence of behavioral, physiological and theoretical evidence presented here provides an initial path and a biologically constrained hypothesis to understand the role of recurrent computations during pattern completion.

**Methods**

An expanded version is presented in the Supplementary Material.

**Psychophysics**. A total of 106 volunteers (62 female, ages 18-34) participated in the behavioral experiments. We performed an experiment with partially visible objects rendered through bubbles (**Fig. 1**) and 3 variations with occluded objects (**Fig. 1, S1**), novel objects (**Fig. 1**, **S8-9**), and stimuli matched to a previous neurophysiological experiment (14) (**Fig. 2**).

**Neurophysiology experiments**. The neurophysiological intracranial field potential data in **Figs. 2** and **3** were taken from reference (14). The neural latency for each image was defined as the time of the peak response in the intracranial field potential and was calculated in single trials (e.g., **Fig. 2C**).

**Computational Models.** We tested state-of-the-art feed-forward vision models, focusing on AlexNet (6) (**Fig. 3**, see **Fig. S4** and **Supplementary Material** for other models), with weights pre-trained on ImageNet (6, 10). As a proof-of-principle, we proposed a recurrent neural network (RNN) model by adding all-to-all recurrent connections to the top feature layer of AlexNet (**Fig. 4A**). The RNN model was defined using only information about the whole objects by setting the recurrent weights based on a Hopfield attractor network (29), as implemented in MATLAB's `newhop` function (30).

**Figure Legends**

**Fig. 1. Backward masking disrupts recognition of partially visible objects**
(**A-B**) Forced-choice categorization task (n=21 subjects). After 500ms of fixation, stimuli were presented for variable exposure times (stimulus onset asynchrony, SOA from 25 to 150 ms), followed by a gray screen (**A**) or a noise mask (**B**) for 500 ms. (**C-D**) Stimuli were either presented unaltered ('Whole'), rendered partially visible ('Partial', **C,** right), or occluded (**D**, right, **Fig. S1**). (**E**) Experimental variation with novel objects (**E, Fig. S8**).
(**F-G**) Behavioral performance as a function visibility for the unmasked (**F**) and masked (**G**) trials. Colors denote different SOAs. Error bars denote SEM. Horizontal line indicates chance level (20%). Bin size = 2.5%. Note the discontinuity in the x-axis to report performance at 100% visibility. (**H**) Average recognition performance as a function of SOA for partial objects (same data replotted from **F-G**, excluding 100% visibility). Performance was significantly degraded by masking (solid gray line) compared to the unmasked trials (dotted gray line) ($p<0.001$, Chi-squared test, d.f.=4). **I**. Performance versus SOA for the occluded stimuli in **D** (note: chance=25% here, see **Fig. S1**). **J**. Performance versus SOA for the novel object stimuli in **E**.

**Fig. 2. The behavioral effect of masking correlated with the neural response latency on an image-by-image basis**
(**A**) Intracranial field potential (IFP) responses from an electrode in the left fusiform gyrus averaged across five categories of whole objects while a subject was performing the task described in **Fig. 1** (no masking, 150 ms presentation time). This electrode showed a stronger response to faces (green). The gray rectangle indicates the stimulus presentation time (150 ms). The shaded area indicates SEM. (see (14) for details).
(**B**) IFP responses for one of the whole objects for the electrode in (**A**) showing single trial responses (gray, n=9) and average response (green). The latency of the peak response is marked on the x-axis.
(**C**) Single-trial responses (n = 1) to 4 partial images of the same object in **B**.

**(D)** A new stimulus set for psychophysics experiments was constructed from the images in 650 trials from two electrodes in the physiology experiments. Raster of the neural responses for the example electrode in **A**, one trial per line, from partial image trials selected for psychophysics. These trials elicited strong physiological responses with a wide distribution of response latencies (sorted by the neural latency). The color indicates the voltage (color scale on bottom). The inset (right) zooms in on the responses to the 82 trials in the preferred category.

**(E)** We measured the effect of backward masking at various SOAs for each of the same partial exemplar images used in the physiology experiment (n=33 subjects) and computed a masking index (MI) for each image (**Methods**). The larger the MI for a given image, the stronger the effect of masking.

**(F)** Correlation between the effect of backward masking (y-axis, MI as defined in **E**) and the neural response latency (x-axis, as defined in **B-C**). Each dot is a single partial object from the preferred category for electrode 1 (blue) or 2 (gray). Error bars for the masking index are based on half split reliability (**Fig. S2**); the neural latency values are based on single trials. There was a significant correlation (Pearson $r = 0.37$, $P = 0.004$, linear regression, permutation test).

**Fig. 3: Standard feed-forward models were not robust to occlusion**

**(A)** Performance of feed-forward computational models (colors) compared to humans (black) (see also **Figs. S4-S5, S9A**). We used the feature representation of a subset of whole objects to train an SVM classifier, and evaluated the model's performance on the feature representation of partial objects (**Methods**). The objects used to train the classifier did not appear as partial objects in the test set. Human performance is shown here (150 ms SOA) for the same set of images. Error bars denote SEM. (5-fold cross-validation).

**(B-C)** The single trial neural latency for each image (**Fig. 2B**) was correlated with the distance of each partial object to its whole object category center for AlexNet pool5 (**B**) and AlexNet fc7 (**C**). Each dot represents a partial object with responses recorded either from electrode 1 (blue dots) or electrode 2 (gray dots). The correlation coefficients and *p* values from the permutation test are shown for each subplot.

**Fig. 4: A dynamic recurrent neural network showed improved performance over time, and was impaired by backward masking**

(**A**) The top-level representation in AlexNet (fc7) receives inputs from fc6, governed by weights $\mathbf{W}_{6 \to 7}$. We added a recurrent loop within the top-level representation (RNN). The weight matrix $\mathbf{W}_h$ governs the temporal evolution of the fc7 representation (**Methods**).

(**B**) Performance of the recurrent neural network $RNN_h$ (blue) as a function of visibility. $RNN_h$ approached human performance (black curve), and represented a significant improvement over the original fc7 layer (red curve). The red and black curves are copied from **Fig. 3A** for comparisons. Error bars = SEM.

(**C**) Temporal evolution of the feature representation for $RNN_h$ as visualized with t-SNE (**Fig. S3B**). Over time, the representation of partial objects approaches the correct category in the clusters of whole images.

(**D**) Overall performance of $RNN_h$ as a function of recurrent time step compared to humans (top dashed line) and chance (bottom dashed line). Error bars = SEM. (5-way cross-validation, **Methods**).

(**E**) Correlation in the classification of each object between $RNN_h$ and humans. Dashed line indicates the upper bound of human-human similarity obtained by computing how well half of the subject pool correlates with the other half. Regressions were computed separately for each category followed by averaging the correlation coefficients across categories. Over time, the model becomes more human-like (**Fig. S6**). Error bars denote S.D. across categories.

(**F**) Effect of backward masking. The same backward mask used in the psychophysics experiments was fed to the $RNN_h$ model at different SOA values (x-axis). Error bars denote SEM (5-way cross-validation). Performance improved with increasing SOA values (**Fig. S10**).

**Funding**

This work was supported by a fellowship from the FITweltweit programme of the German Academic Exchange Service (DAAD) (MS), NSF STC award CCF-123121 (GK) and NIH award R01EY026025 (GK).

**Acknowledgments**

We thank Carlos Ponce, Alan Yuille, Siamak Sorooshyari and Guy Ben-Yosef for useful discussions and comments.

**Data availability statement**

All data and code (including image databases, behavioral measurements, physiological measurements and computational algorithms) will be made available upon publication through the lab's web site and GitHub repository (http://klab.tch.harvard.edu/resources/Tangetal_RecurrentComputations.html).

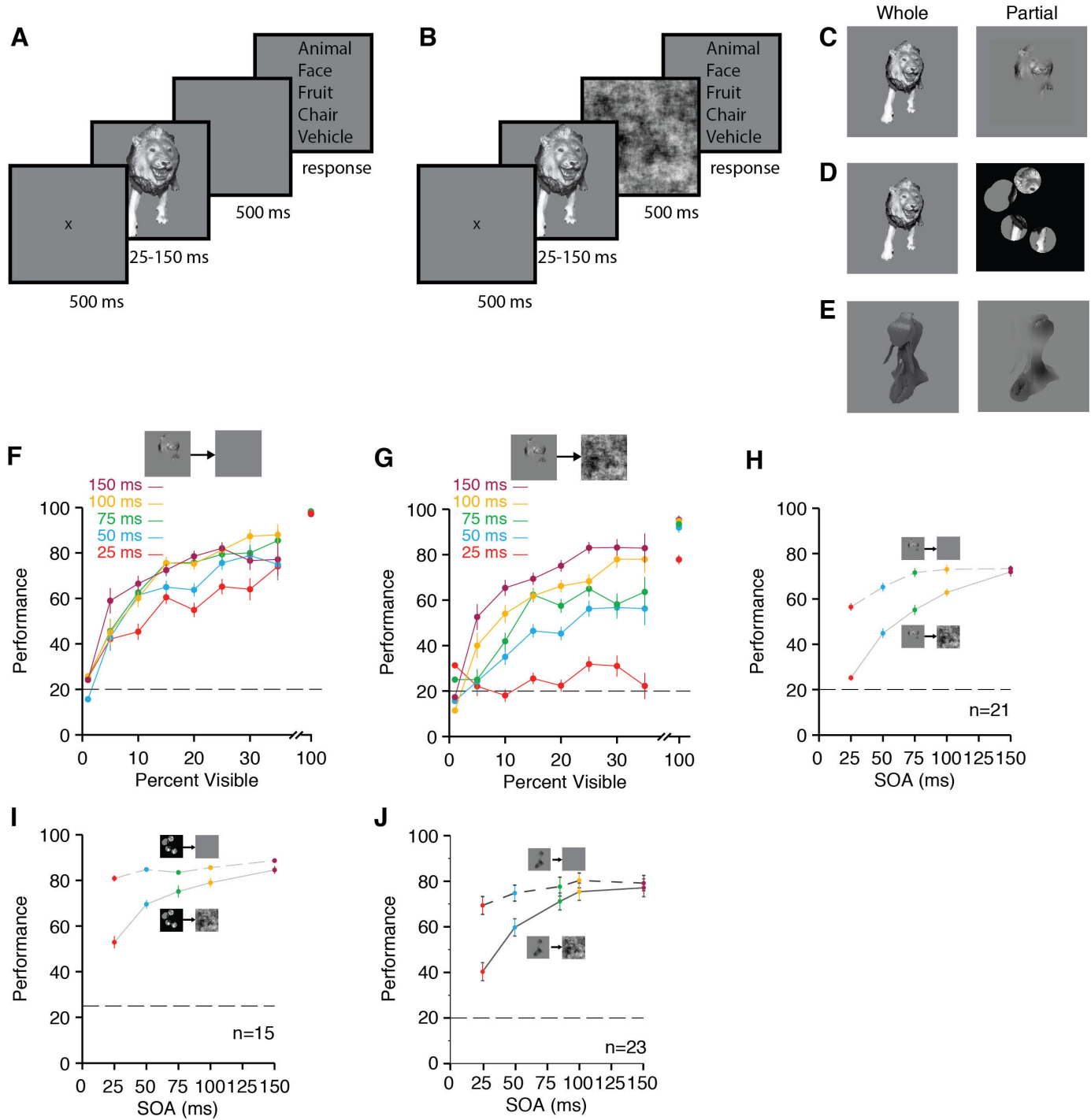

Figure 2

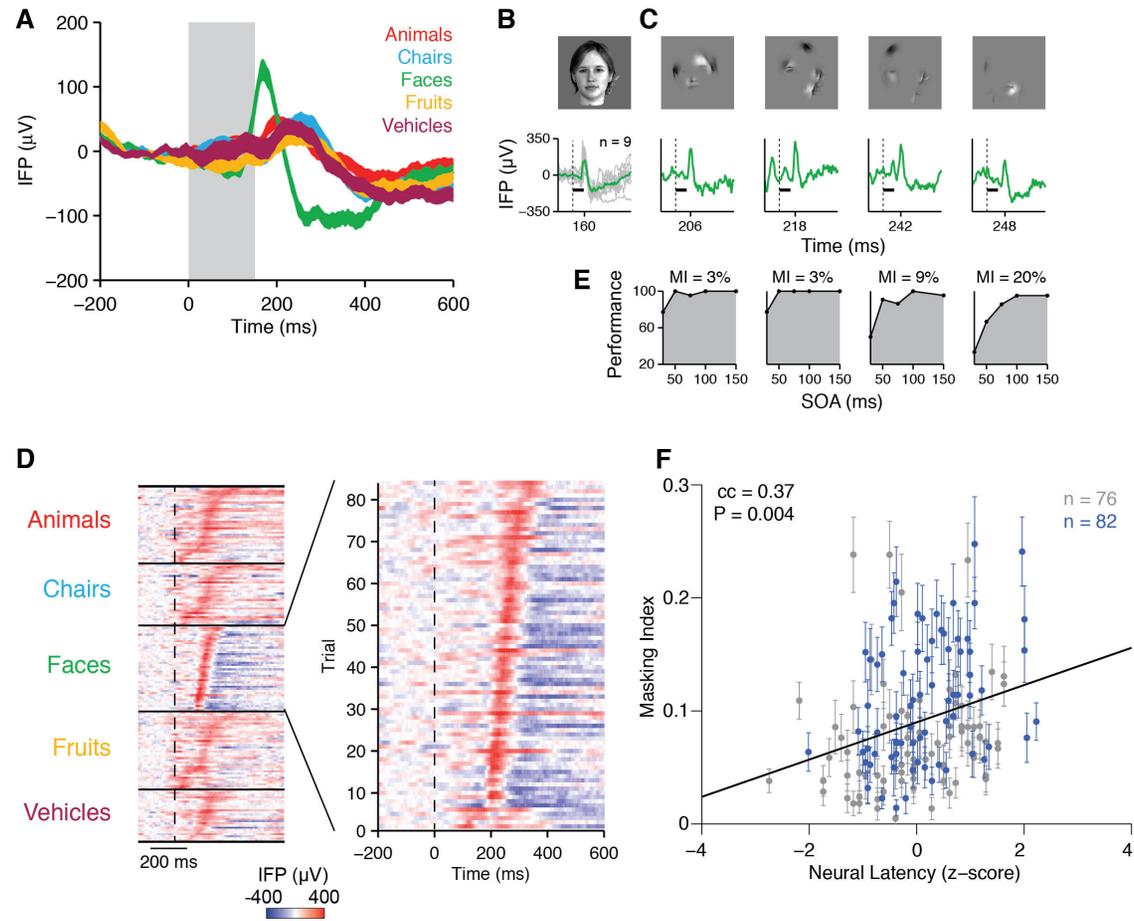

Figure 3

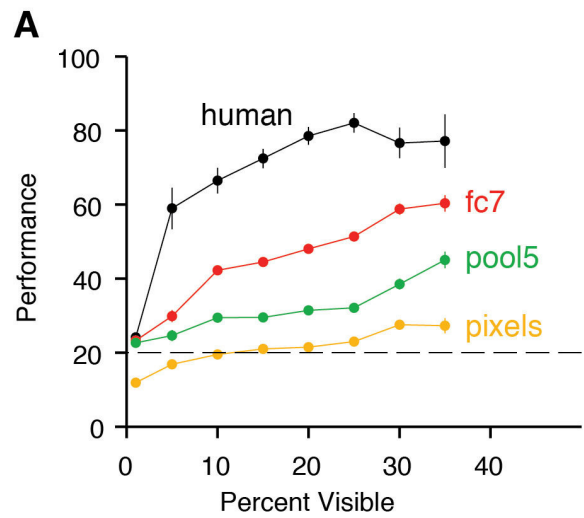
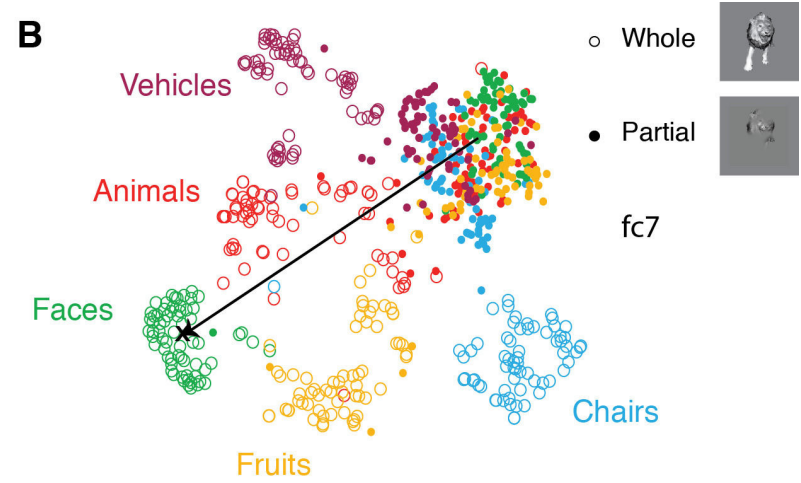
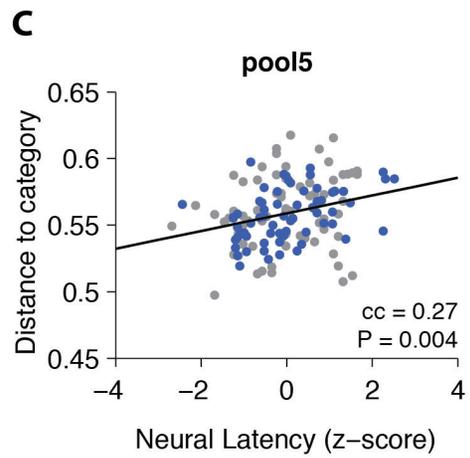
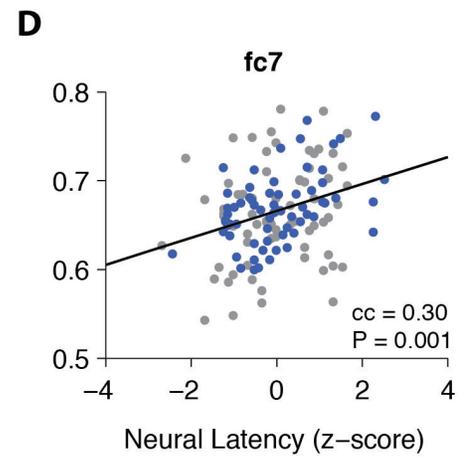

Figure 4

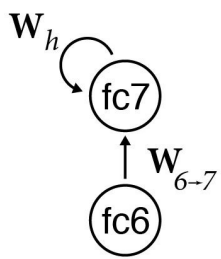

**A**

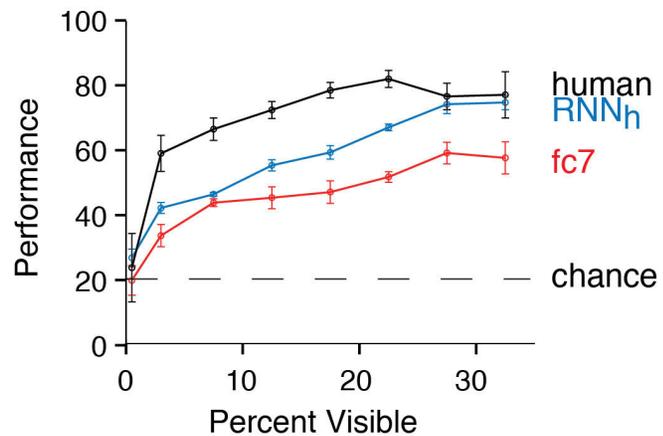

**B**

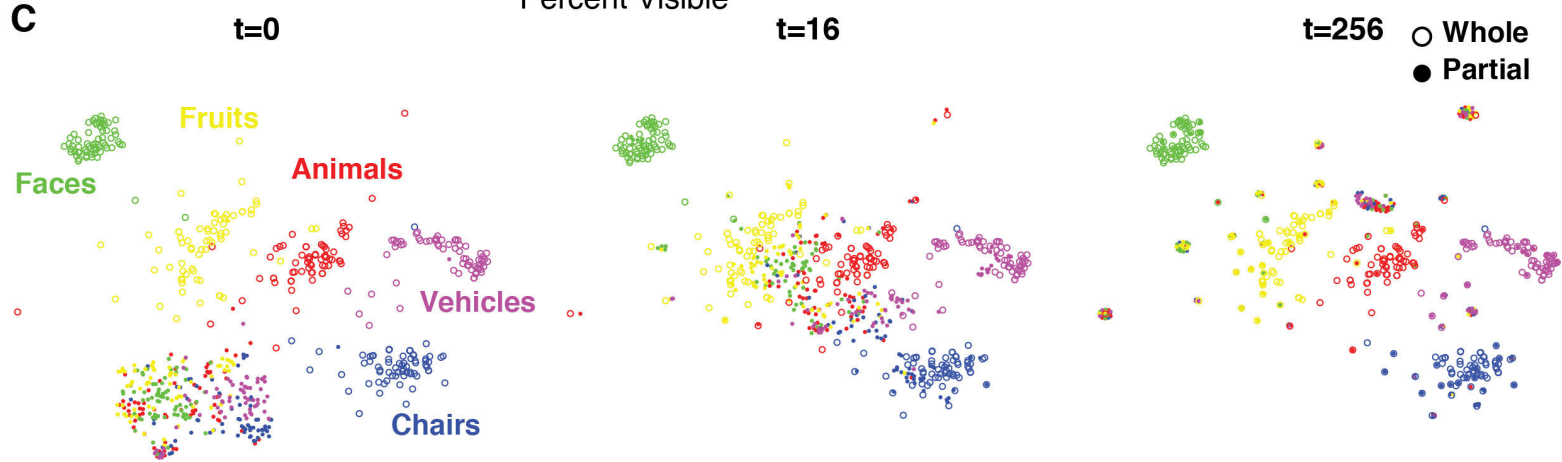

**C** t=0, t=16, t=256 — Whole (○) / Partial (●); Faces, Fruits, Animals, Vehicles, Chairs

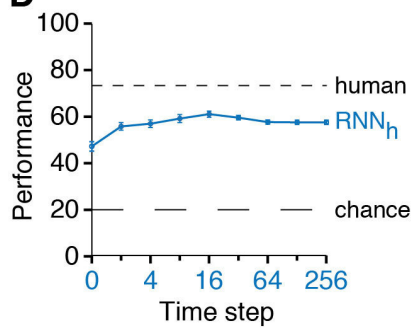

**D**

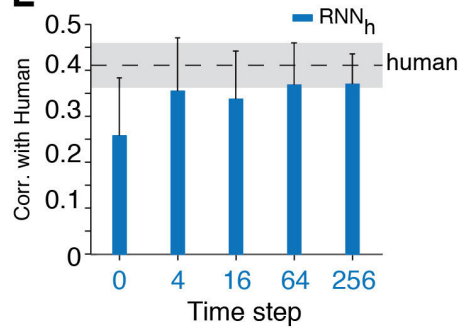

**E**

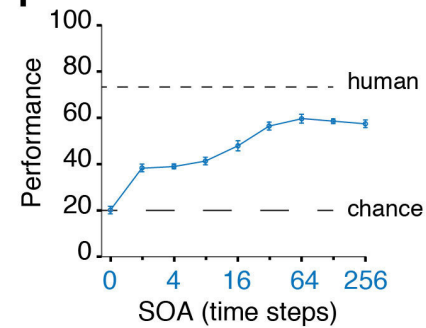

**F**

# Supplementary Figure 1

**A**

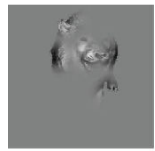

Partial

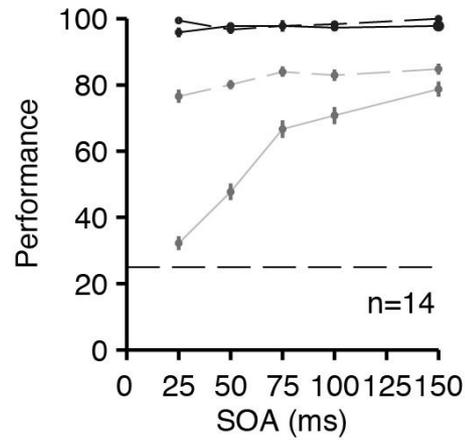

**B**

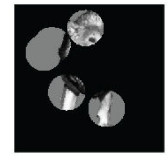

Occluded

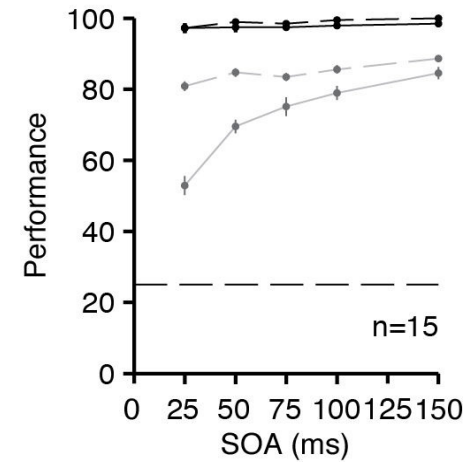

**Figure S1: Robust performance with occluded stimuli**

We measured categorization performance with masking (solid lines) or without masking (dashed lines) for (**A**) partial and (**B**) occluded stimuli on a set of 16 exemplars belonging to 4 categories (chance = 25%, dashed lines). There was no overlap between the 14 subjects that participated in (**A**) and the 15 subjects that participated in (**B**). The effect of backward masking was consistent across both types of stimuli. The black lines indicate whole objects and the gray lines indicate the partial and occluded objects. Error bars denote SEM.

# Supplementary Figure 2

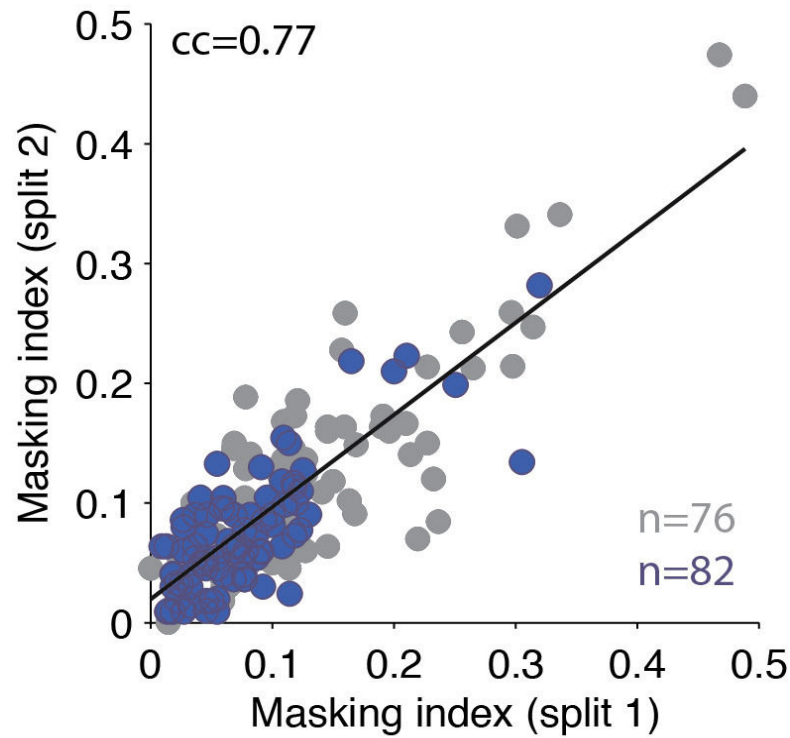

**Figure S2: Example half-split reliability of psychophysics data**

**Figure 2E** in the main text reports the masking index, a measure of how much recognition of each individual image is affected by backward masking. This measure is computed by averaging performance across subjects. In order to evaluate the variability in this metric, we randomly split the data into two halves and computed the masking index for each image for each half of the data. This figure shows one such split and how well one split correlates with the other split. **Figure 2F** shows error bars defined by computing standard deviations of the masking indices from 100 such random splits.

# Supplementary Figure 3

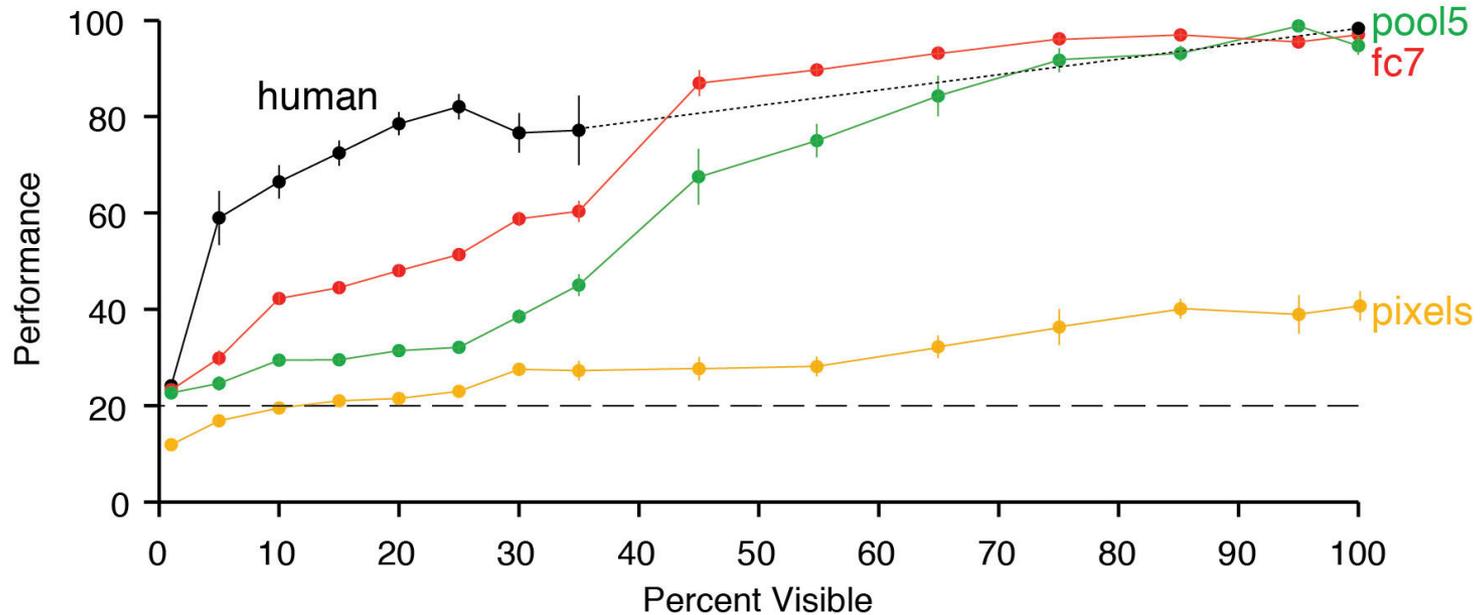

**Figure S3: Bottom-up models can recognize minimally occluded images**

Extension to **Fig. 3A** showing that bottom-up models successfully recognize objects when more information is available (**Fig. 3A** showed visibility values up to 35% whereas this figure extends visibility up to 100%). The format and conventions are the same as those in **Fig. 3A**. The black dotted line shows interpolated human performance between the psychophysics experimental values measured at 35% and 100% visibility levels.

# Supplementary Figure 4

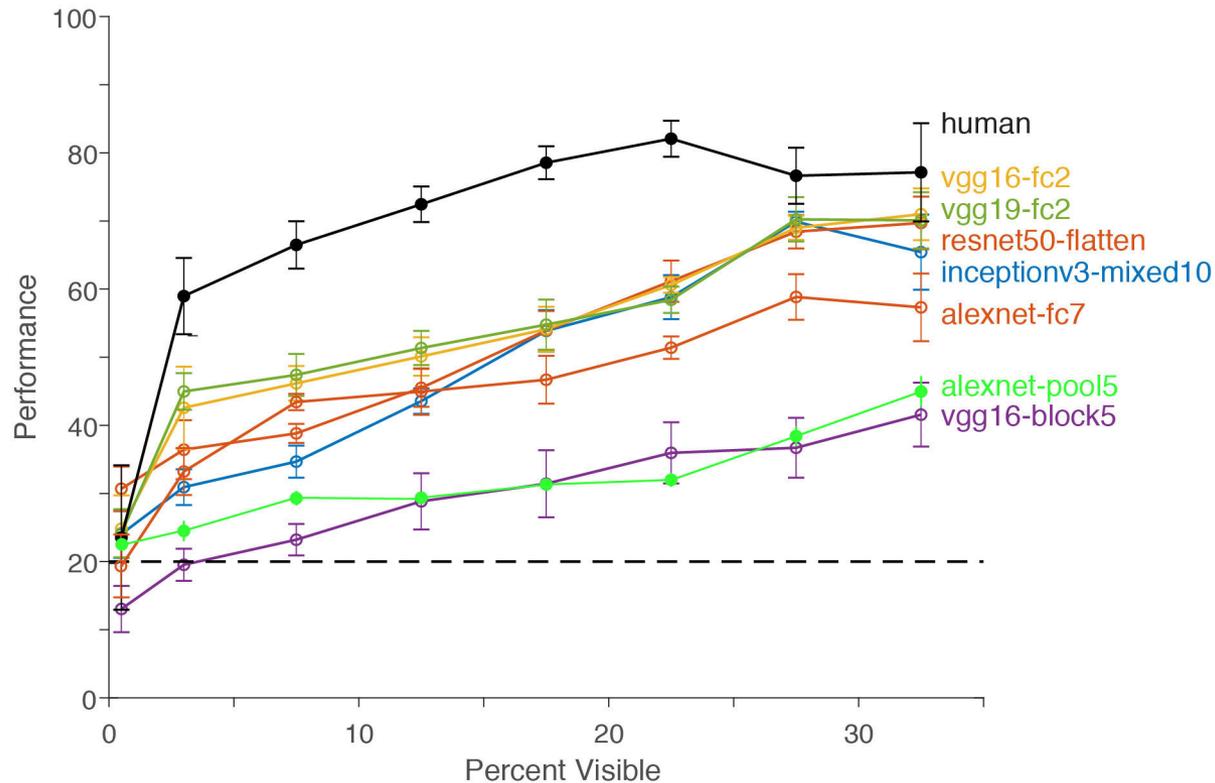

**Figure S4: All of the purely feed-forward models tested were impaired under low visibility conditions**

The human, AlexNet-pool5 and AlexNet-fc curves are the same ones shown in **Figure 3A** and are reproduced here for comparison purposes. This figure shows performance for several other models: VGG16-fc2, VGG19-fc2, ResNet50-flatten, inceptionV3-mixed10, VGG16-block5 (see text for references). In all cases, these models were pre-trained to optimize performance under ImageNet 2012 and there was no additional training (see also **Figure S5** for fine tuning results). An expanded version of this figure with many other layers and models can be found on our web site:
http://klab.tch.harvard.edu/resources/Tangetal_RecurrentComputations.html

# Supplementary Figure 5

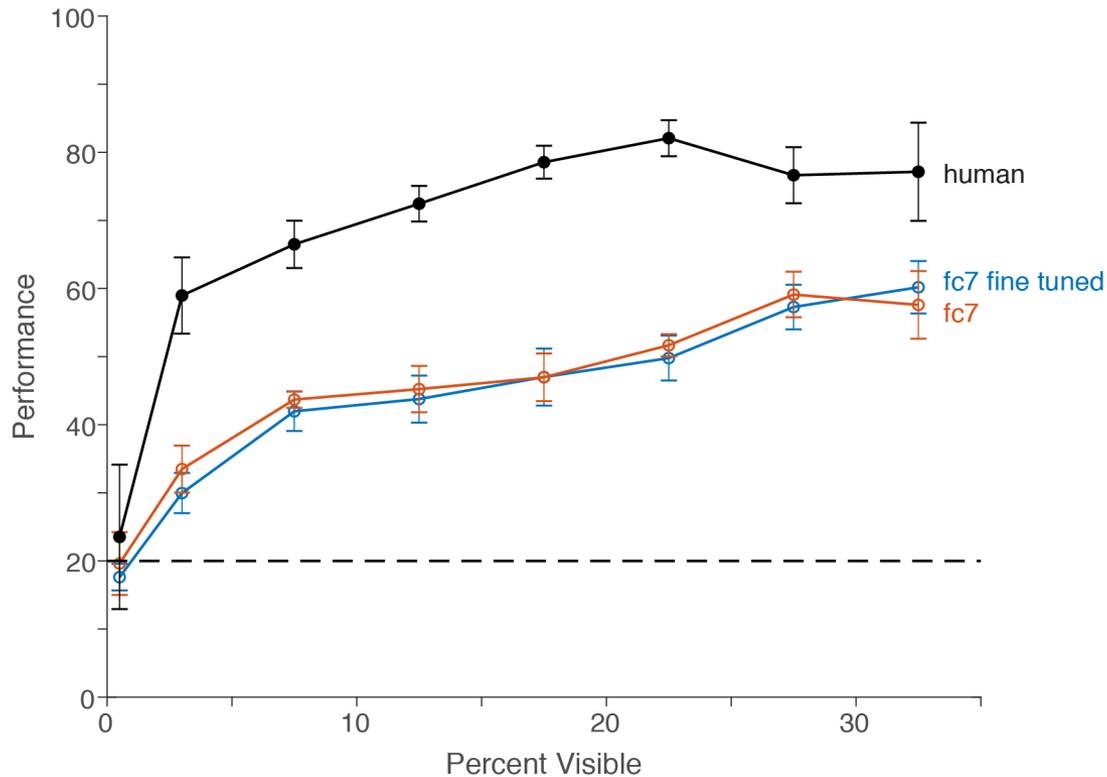

**Figure S5: Fine-tuning did not improve performance under heavy occlusion**

The human and fc7 curves are the same ones shown in **Figure 3A** and are reproduced here for comparison purposes. The pre-trained AlexNet network used in the text was fine tuned using back-propagation with the set of *whole* images from the psychophysics experiment (in contrast with the pre-trained Alexnet network which was trained using the Imagenet 2012 data set). The fine-tuning involved all layers (**Methods**).

# Supplementary Figure 6

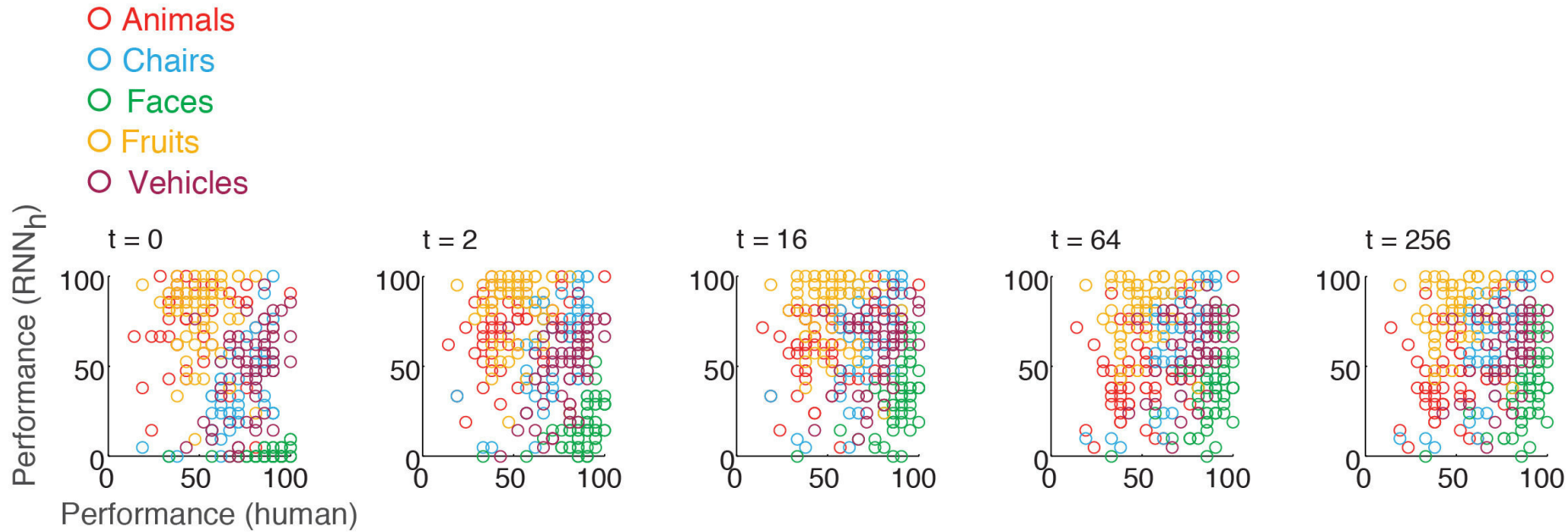

**Figure S6: Correlation between RNN$_h$ model and human performance for individual objects as a function of time**

At each time step in the recurrent neural network model (RNN$_h$), the scatter plots show the relationship between the model's performance on individual partial exemplar objects and human performance. Each dot is an individual exemplar object. In **Fig. 4E** we report the average correlation coefficient across all categories.

# Supplementary Figure 7

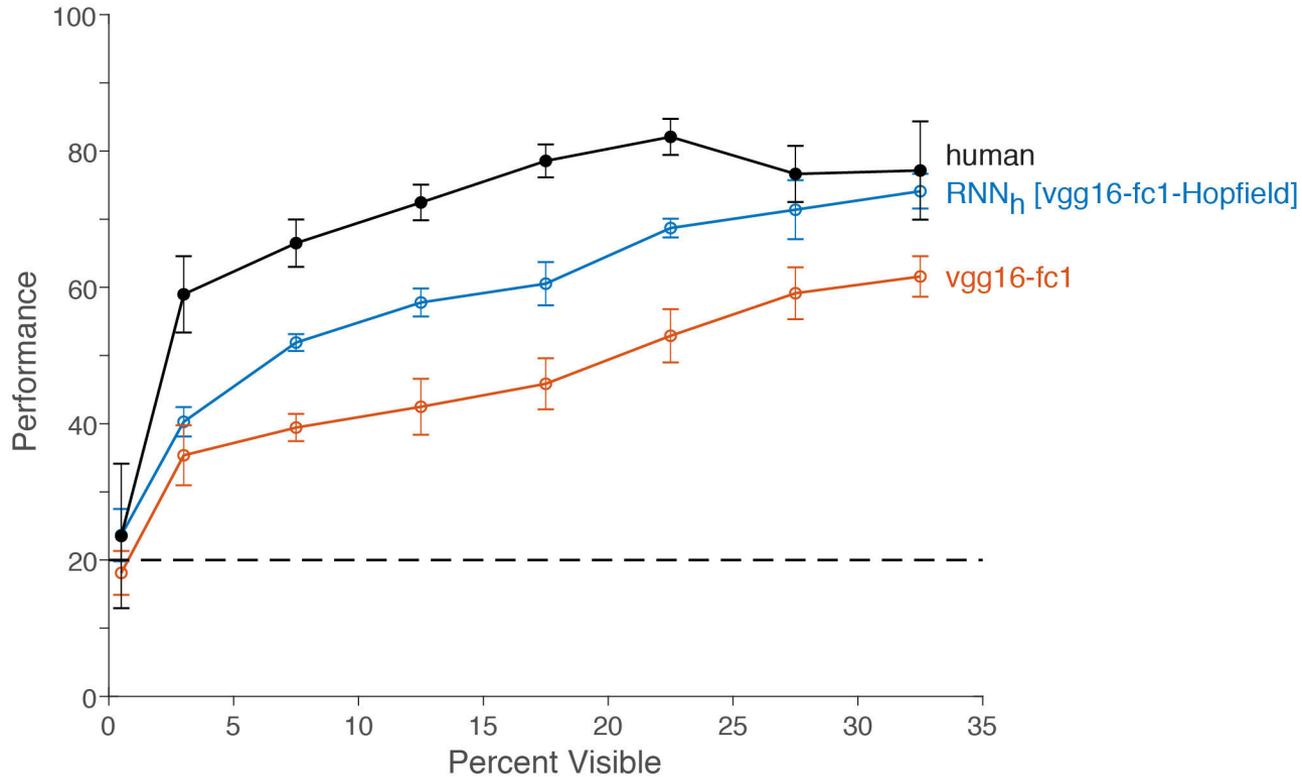

**Figure S7: Adding recurrent connectivity to VGG16 also improved performance**

This Figure parallels the results shown in **Figure 4B** for AlexNet, here using the VGG16 network, implemented in keras (**Methods**). The results shown here are based on using 4096 units from the fc1 layer. The red curve (vgg16-fc1) corresponds to the original model without any recurrent connections. The implementation of the $RNN_h$ model here (VGG16-fc1-Hopfield) is similar to the one in **Figure 4B**, except that here we use the VGG16 fc1 activations instead of the AlexNet fc7 activations. An expanded version of this figure with similar results for several other layers and models can be found on our web site: http://klab.tch.harvard.edu/resources/Tangetal_RecurrentComputations.html

# Supplementary Figure 8

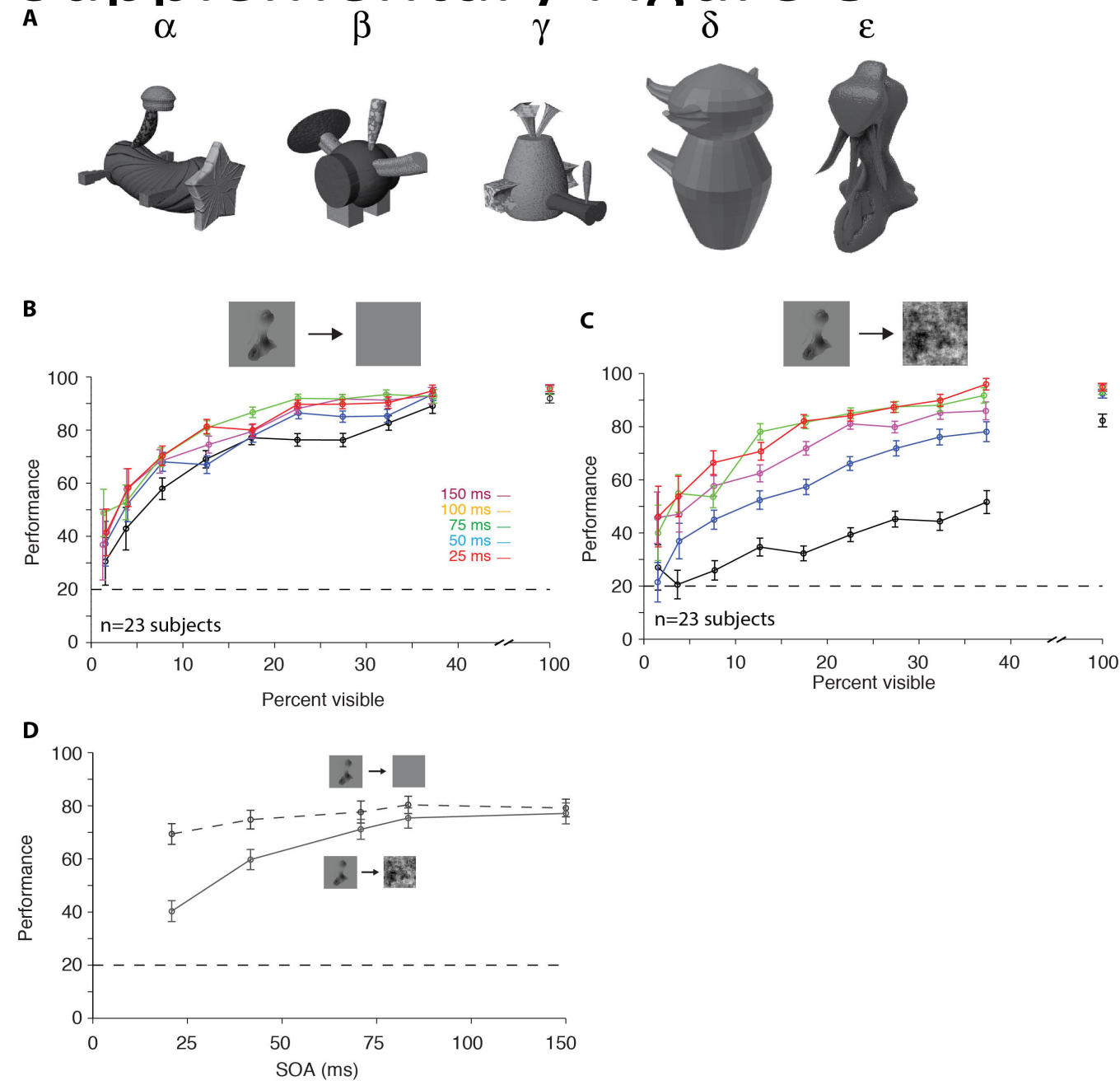

**Figure S8: Robust recognition of *novel* objects under low visibility conditions**

**A**. Single exemplar from each of the 5 novel object categories (**Methods**).

(**B-C**) Behavioral performance for the unmasked (**B**) and masked (**C**) trials. The experiment was identical to the one in **Figure 1** and the format of this figure follows that in **Figure 1F-G**. The colors denote different SOAs. Error bars=SEM. Dashed line = chance level (20%). Bin size=2.5%. Note the discontinuity in the x-axis to report performance for whole objects (100% visibility). (**D**) Average recognition performance as a function of the stimulus onset asynchrony (SOA) for partial objects (same data and conventions as **B-C**, excluding 100% visibility). Error bars=SEM. Performance was significantly degraded by masking (solid) compared to the unmasked trials (dotted) ($p<0.0001$, Chi-squared test, d.f.=4).

# Supplementary Figure 9

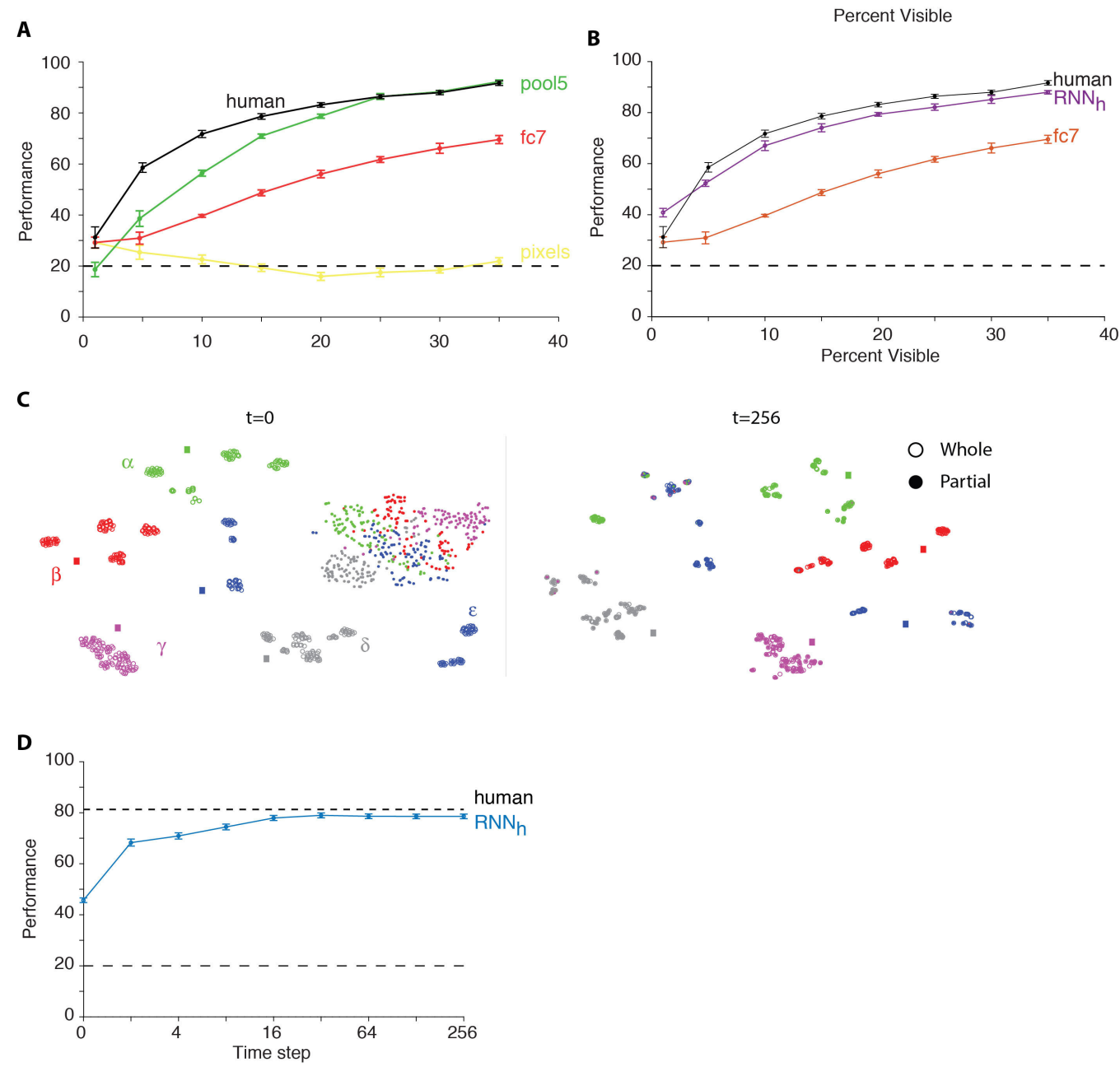

**Figure S9: The performance of feed-forward and recurrent computational models for *novel* objects was similar to those for known object categories**

**A.** Performance of feed-forward computational models (format as in **Figure 3A**) for novel objects.

**B**. Performance of the recurrent neural network $RNN_h$ (format as in **Figure 4B**) for novel objects.

**C**. Temporal evolution of the feature representation for $RNN_h$ (format as in **Figure 4C**). The colors and greek letters denote the five object categories (see examples in **Figure S8A**).

**D.** Performance of $RNN_h$ as a functon of recurrent time for novel objects (format as in **Figure 4D**).

# Supplementary Figure 10

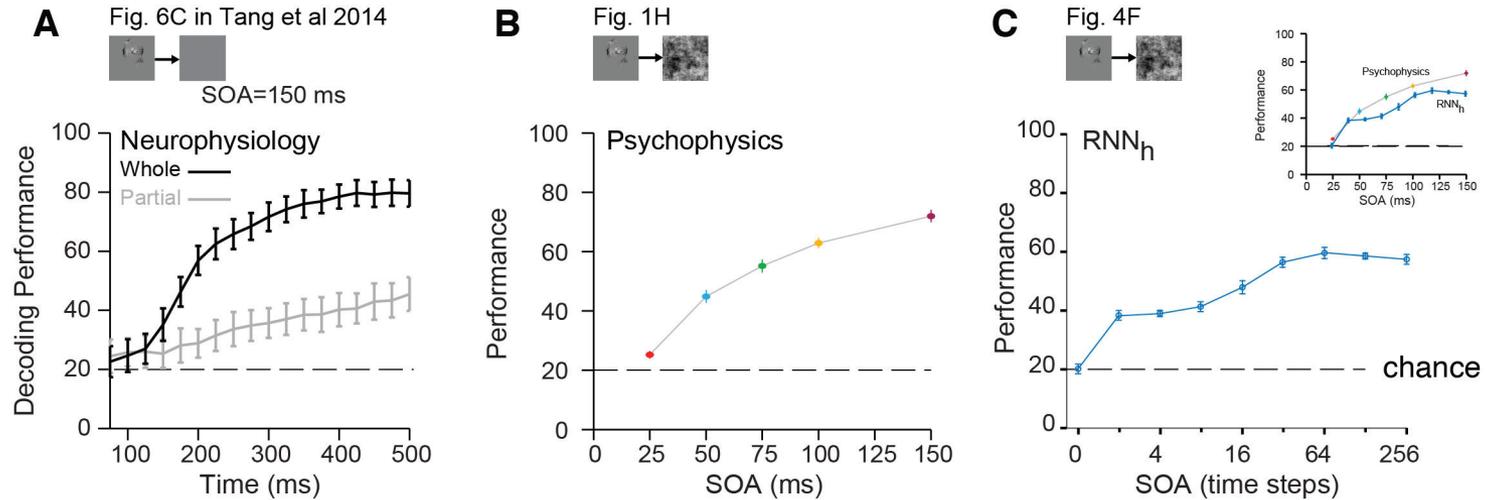

**Figure S10: Side-by-side comparison of neurophysiological signals, psychophysics and computational model**
**A.** Reproduction of Figure 6C from Tang et al 2014. This figure shows the dynamics of decoding object information for whole objects and (black) and partial objects (gray) from neurophysiological recordings as a function of time post stimulus onset (see Tang et al 2014 for details.
**B**. Reproduction of **Figure 1H** (behavior).
**C**. Reproduction of **Figure 4F** ($RNN_h$ model).
Above each subplot, the experiment schematic highlights that **A** involves no masking and fixed SOA = 150 ms whereas **B, C** involve masking and variable SOAs. The inset in part **C** directly overlays the results of the $RNN_h$ model in **C** onto the results of the psychophysics experiment in **B**. In order to create this plot, we mapped 0 time steps to 25ms, 256 time steps to 150 ms and linearly interpolated the time steps in between.

# Supplementary Figure 11

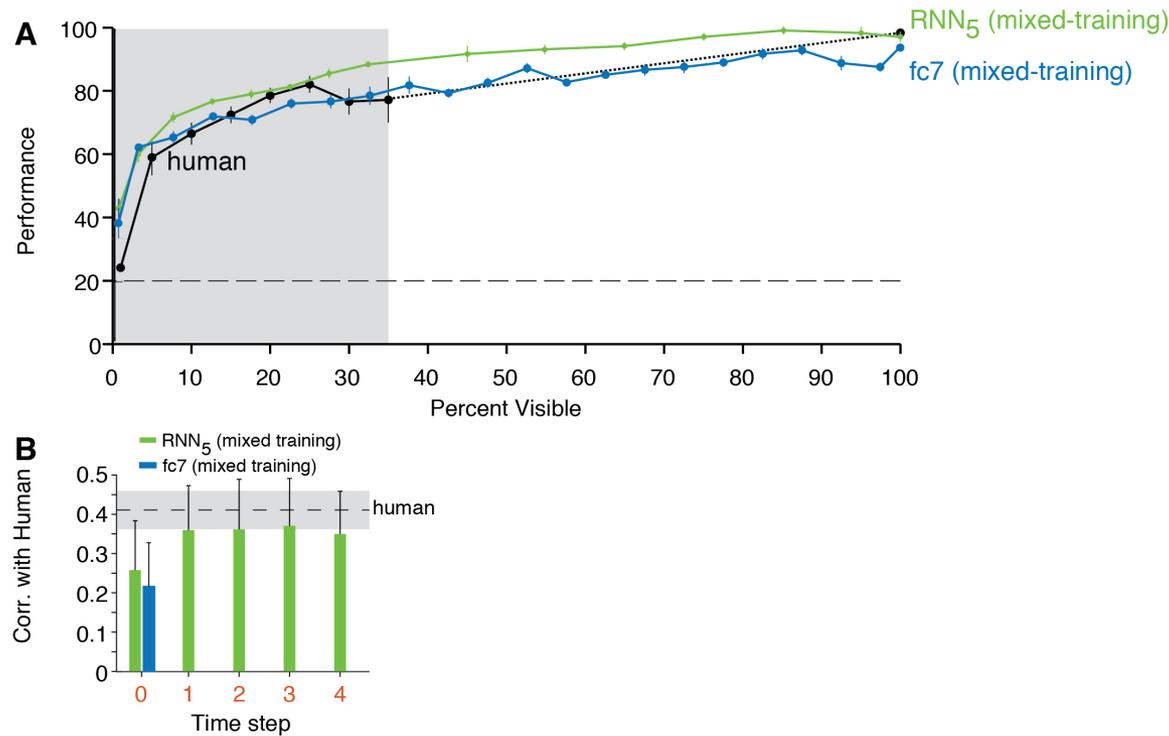

**Figure S11: Mixed training regimes**.

**A.** This figure follows the format of **Fig3A**, **4B** and **S3A, S4, S5, S7, S9A-B**. The black line shows human performance and is copied from **Fig. 3A** for comparison purposes. The green and blue lines show the recurrent model (RNN$_5$) and bottom-up model (AlexNet fc7), respectively, trained in a mixed regime that included the occluded objects with visibility levels within the gray rectangle (the same ones used to evaluate human psychophysics performance). As noted in the text, we emphasize that this figure involves a different training regime from the ones in the previous figures and therefore one cannot directly compare performance with the previous figures.

**B.** This figure follows the format of **Fig. 4E**. The green and blue bars show the correlation between human and model for the recurrent model and bottom-up model, respectively, both trained using occluded objects. The gray rectangle shows human-human correlation, see **Fig. 4E** for details..